\documentclass[amsmath,
amssymb,
a4paper,
aip, % society option (aps, aip)
jcp, % society's journal style (prl, apl, ..)
reprint, 
twocolumn, %reprint, % layout option (preprint, reprint, twocolumn)
fleqn,
showpacs,
floatfix]{revtex4-2}
\usepackage{amsmath}
\usepackage[table]{xcolor}
\usepackage[T1]{fontenc}
\usepackage{lmodern}
\usepackage{bbm}
\usepackage{microtype}
\usepackage{mathtools}%, cuted}
\usepackage{mathcomp}
\usepackage{color}
\usepackage{tabularx}
\usepackage{stmaryrd}
\usepackage{booktabs}
\usepackage{multirow}
\usepackage{xcolor}
\usepackage[noend]{algorithm2e}
\usepackage[
format=hang,
indention=-1.0cm,
justification=RaggedRight,
font=small
]{caption}
\usepackage{graphicx} % Include figure files
\usepackage{dcolumn} % Align table columns on decimal point
\usepackage{bm} % bold math
\usepackage[version=4]{mhchem}
\usepackage{stackrel}
\usepackage{verbatim}   % useful for program listings
\usepackage{subfigure}  % use for side-by-side figures
\usepackage{hyperref}   % use for hypertext links, including those to external documents and URLs
\usepackage{float}
\usepackage{amsfonts} %helps out with particular math fonts, e.g. Z and R
\usepackage[geometry]{ifsym}
\usepackage[utf8]{inputenc}
\usepackage[english]{babel}
\usepackage{calc}
\usepackage{graphicx} 
\usepackage{cancel}
\graphicspath{{}} 
\newlength\myheight
\newlength\mydepth
\settototalheight\myheight{Xygp}
\newcommand{\be}{\begin{equation}}
\newcommand{\ee}{\end{equation}}
\newcommand{\bea}{\begin{eqnarray}}
\newcommand{\eea}{\end{eqnarray}}

\newcommand{\mathleft}{\@fleqntrue\@mathmargin0pt}
\newcommand{\mathcenter}{\@fleqnfalse}
\usepackage{academicons}
\definecolor{orcidlogocol}{HTML}{A6CE39}

\usepackage{orcidlink}
\makeatletter
\renewcommand{\@algocf@capt@plain}{above}%
\makeatother

\begin{document}

\title{On mechanistically accessible copolymer sequences}
\author{Richard Golnik\orcidlink{0000-0002-8582-5006}}
\affiliation{Universität Leipzig, Germany} 

%\affiliation{Bioinformatics Group, Department of Computer Science, Leipzig University,Härtelstraße 16–18, D-04107 Leipzig, Germany} 

\author{Nicola Vassena\orcidlink{0000-0001-5411-4976} }
\affiliation{Universität Leipzig, Germany}

\author{Alex Blokhuis\orcidlink{0000-0002-4594-596X}}
\email{alexander.willem@imdea.org}
\affiliation{Instituto IMDEA Nanociencia, Calle Faraday 9, 28049 Madrid, Spain}

%\date{February 2026}

\begin{abstract}
%Traditionally, copolymerization is often treated as a process in which monomer incorporation depends at most on nearest neighboring incorporated monomers. However, non-nearest neighbor interactions can be readily realized and nontrivial periodic self-assembled sequences have already been experimentally observed. 
Structural observations in polymer chemistry are often rationalized through underlying mechanistic insights. The study of families of mechanisms - and the phenomenology and lawlike behavior that can accompany them - is less common. Here we construct a theory that links a mechanistic feature - the number of nearest neighbors $k$ that influence monomer incorporation - with the accessible copolymer sequence space. 

We find that the reaction mechanisms can be mapped 1-to-1 to a regular de Bruijn graph of degree $k$. We calculate the periodic copolymer sequences as function of neighbor index $k$, and number of monomers $s$. We show that the number of periodic sequences grows explosively with $s,k$, bounded from below by $s!^{s^{k-1}}/s^k$. We find that periodic k-polymerization is governed by several simple laws describing allowed sequences, number of specific reactions, repeating unit length and the allowed and disallowed sequence coexistence in block copolymers. 

We formulate Simple Assembly Rules (SAR) that capture commonly observed behavior and design of  literature systems, and characterize the possible k-polymerizations they are allowed to access ($k\leq2$) as function of symmetries and illustrate our findings with reported examples from the coordination chemistry and supramolecular chemistry literature. 

We discuss chemical-structural contributions by which nontrivial ($k>1$) sequences are formed, highlighting the importance steric hindrance, long-range interactions and degrees of freedom.

\end{abstract}

\maketitle

\section{Introduction}

New structural observations in polymer chemistry have often given rise to new mechanistic hypotheses. Polymers with blocks of different tacticities were rationalized in terms of multiple states of the polymer tip\cite{Coleman1963} (Coleman-Fox mechanism), with distinct reactivities. Similarly, an 'oscillating polymerization catalyst'\cite{coates1995}, which interconverts between different states (which may bias different tacticities). A minor amount of chiral monomer (sergeant) leading to a disproportionate amount of one helical polymer was rationalized through two helical polymer states with one biased by the added monomer (sergeants-and-soldiers mechanism), existing either as distinct polymers or as a block copolymer. Interestingly, this amplification phenomenon does not require monomers to be sensitive to the polymer tip (neighbor index $k = 0$). 

Whereas individual phenomenology is understood in terms of associated models, a broader overarching theory linking mechanisms to e.g. sequence appears to remain an open challenge. The possible space of repeating copolymer sequences was characterized by Balaban and Artemi in 1987-1988\cite{Balaban1987_I,Artemi1987_II,Balaban1988}, in terms of possible repeating unit as function of length. Although an explosive diversity of sequences can exist (and is increasingly realized), canonical polymerization models currently capture only the simplest sequences. 

Understanding the formation of more complex sequences is pertinent both towards its rational design and towards recognizing its phenomenology where it is already happening, but has heretofore been overlooked. To this end, we here introduce the notion of k-polymerization, where monomer additions depends on k terminal monomer units in a polymer. 

We will first formalize k-polymerization and motivate its use both for irreversible polymerization and for polymerization equilibrium. We in particular consider conditions where reactions become specific and give rise to periodic sequences. These conditions give rise to several laws governing the possible sequences that can form for k-polymerization, as well as which sequences can coexist in block-copolymers. We characterize the sequences that can form as function of $k$, and show that this diversity grows explosively. 

As a first step towards understanding the microscopic origins of k-polymerization, we formulate Simple Assembly Rules (SAR), which describe a large class of reported systems in the literature. We show that these can reach values of $k$ of at most 2. Whether this cap is attained is governed by orientational symmetries among monomers and captured by index laws. We apply our findings to characterize a large number of reported synthetic examples in the literature.

%An explosive diversity of sequences can exist and are increasingly realized. 

%Most models can be said to have been motivated by a specific problem. What - to our knowledge - remains outstanding, is a broader view that starts from various families of models to compare the phenomenology they can give rise to, such as the sequ

%Whereas many models have been 

%Whereas 

%The reverse 

%Structure informs reactivity and mechanism. 

%Sergeant-and-soldiers model

%Mechanistic detail 

\section{k-copolymerization}

\begin{figure}[tbhp!]
\centering
\includegraphics[width=0.9\linewidth]{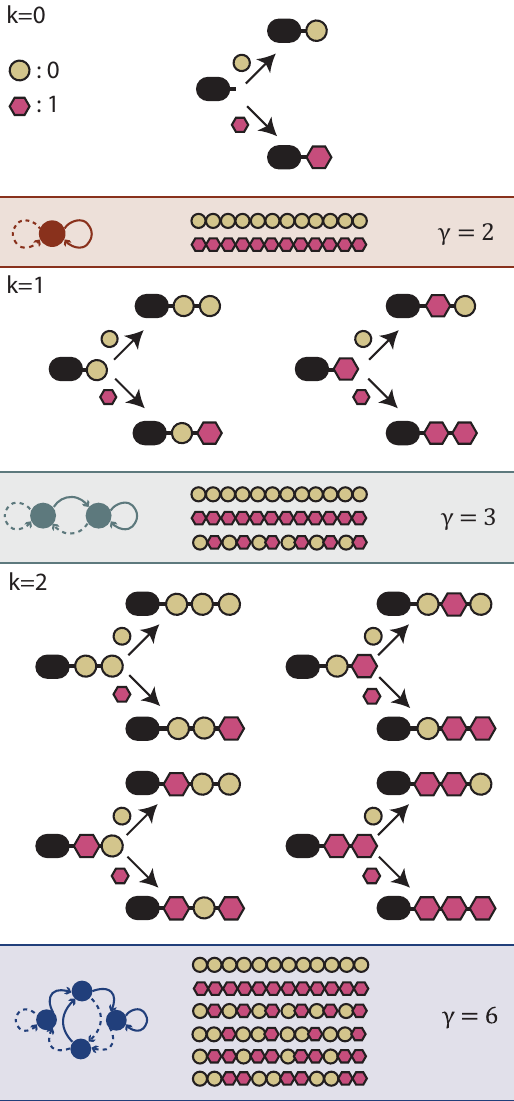}
\caption{k-copolymerization for s=2 monomer types '0', '1' for k=0,1,2. A corresponding regular de Bruijn graph is depicted for each k (see Fig. \ref{fig:debruijn}), along wiht the attainable periodic sequences, whose number is denoted by $\gamma$.}
\label{fig:elongations}
\end{figure}

We define k-copolymerization (Fig. \ref{fig:elongations}) as the polymerization of $s$ monomer types $\ce{A}, \ce{B}, .. , \ce{S}$, dependent on the last $k$ incorporated monomers, hereafter 'vicinality' $k$. For each elongation step, we thus need to know the monomer sequence of the polymer tip, for which there exist $s^k$ states, each a priori admitting $s$ reactions, for a total of $s^{k+1}$ reactions.

\begin{figure}[tbhp!]
\centering
\includegraphics[width=0.9\linewidth]{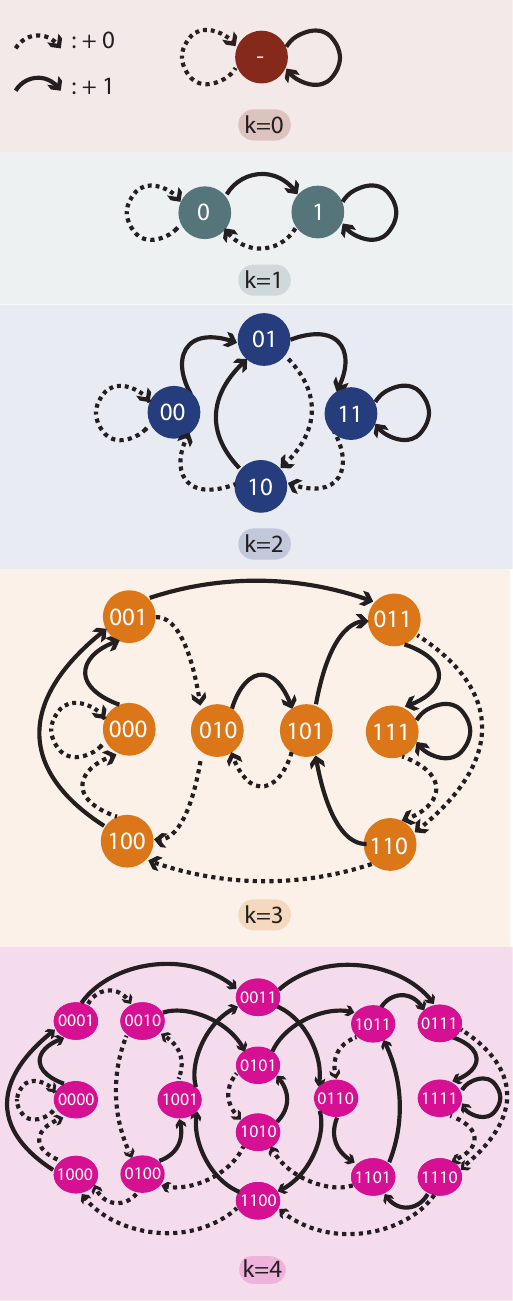}
\caption{Successive regular de Bruijn Graphs (k=0 to 4) for $s=2$ monomers ('0', '1'). Each periodic sequence for k-copolymerization corresponds to an elementary circuit in the de Bruijn graph of order k.}
\label{fig:debruijn}
\end{figure}

Let us denote $P$ a (co)polymer. For 0-copolymerization ($k=0$, $s=2$), there is only 1 tip state, and $2$ reactions
\bea
\ce{P} + \ce{A} \underset{1}{\leftrightarrows} \ce{P A} \ \ \ \ce{P} + \ce{B} \underset{2}{\leftrightarrows} \ce{P B} \nonumber
\eea
for 1-copolymerization ($s=2$) we need to specify the monomer at the tip
\bea
\ce{PA} + \ce{A} \underset{1}{\leftrightarrows} \ce{PAA} \ \ \ \ce{PA} + \ce{B} \underset{2}{\leftrightarrows} \ce{PAB} \ \ \ \nonumber \\
\ce{PB} + \ce{A} \underset{3}{\leftrightarrows} \ce{PBA} \ \ \ \ce{PB} + \ce{B} \underset{4}{\leftrightarrows} \ce{PBB} \ \ \  \nonumber
\eea
for 2-copolymerization (k=2, s=2) we need to specify the last two monomers at the tip
\bea
\! \! \! \! \! \! \ce{PAA} + \ce{A} \underset{1}{\leftrightarrows} \ce{PAAA} \ \ \ \ce{PAA} + \ce{B} \underset{2}{\leftrightarrows} \ce{PAAB} \ \ \ \ \ \ \ \ \ \  \nonumber \\
\! \! \! \! \! \! \ce{PAB} + \ce{A} \underset{3}{\leftrightarrows} \ce{PABA} \ \ \ \ce{PAB} + \ce{B} \underset{4}{\leftrightarrows} \ce{PABB} \ \ \ \ \ \ \ \ \ \ \ \nonumber \\
\! \! \! \! \! \! \ce{PBA} + \ce{A} \underset{5}{\leftrightarrows} \ce{PBAA} \ \ \ \ce{PBA} + \ce{B} \underset{6}{\leftrightarrows} \ce{PBAB} \ \ \ \ \ \ \ \ \ \  \nonumber \\
\! \! \! \! \! \! \ce{PBB} + \ce{A} \underset{7}{\leftrightarrows} \ce{PBBA} \ \ \ \ce{PBB} + \ce{B} \underset{8}{\leftrightarrows} \ce{PBBB} \ \ \ \ \ \ \ \ \ \ \nonumber  
\eea
incorporation depending on the last $k$ tip monomers for $s$ monomer types is thus described in terms of $s^{k+1}$ reactions. The system of tip sequences and reactions (Fig. \ref{fig:elongations}) can equivalently be represented by a regular de Bruijn graph, as shown in Fig. \ref{fig:debruijn}. 

Our inquiry focuses on periodic sequences deterministically attainable by selectively favoring certain elongation reactions. % We do not require these sequences to be defect-free, but 

\subsection{Kinetic and thermodynamic descriptions}

The two most common contexts for the study of polymerization models are i) irreversible polymerization and ii) thermodynamic equilibrium. On the level of a single polymer chain, we can write the former as a Markov process with a master equation of the form
\be
\pmb{p}_{n+1} = T_{k,s} \pmb{p}_{n} 
\ee
where $\pmb{p}_n$ is a vector of occupation probabilities for each tip state with a polymer of preceding length $n$ and $T$ a transfer matrix containing transition probabilities among tip states upon addition of one monomer. For $k=2, s=2$, we have 
\bea
\pmb{p}_n &=& (p_{\ce{X_nAA}}, p_{\ce{X_nBA}}, p_{\ce{X_nBA}}, p_{\ce{X_nBA}} )^T \nonumber \\
T_{2,2} &=& \begin{pmatrix}
\pi^{\ce{AA}}_{\ce{AA}} & \pi^{\ce{AA}}_{\ce{BA}} & \pi^{\ce{AA}}_{\ce{AB}} & \pi^{\ce{AA}}_{\ce{BB}} \\ 
\pi^{\ce{BA}}_{\ce{AA}} & \pi^{\ce{BA}}_{\ce{BA}} & \pi^{\ce{BA}}_{\ce{AB}} & \pi^{\ce{BA}}_{\ce{BB}} \\ \pi^{\ce{AB}}_{\ce{AA}} & \pi^{\ce{AB}}_{\ce{BA}} & \pi^{\ce{AB}}_{\ce{AB}} & \pi^{\ce{AB}}_{\ce{BB}} \\ \pi^{\ce{BB}}_{\ce{AA}} & \pi^{\ce{BB}}_{\ce{BA}} & \pi^{\ce{BB}}_{\ce{AB}} & \pi^{\ce{BB}}_{\ce{BB}}   \end{pmatrix} \\
&=& \begin{pmatrix} \pi^{\ce{AA}}_{\ce{AA}} & \pi^{\ce{AA}}_{\ce{BA}} & 0 & 0 \\ 
0 & 0 & \pi^{\ce{BA}}_{\ce{AB}} & \pi^{\ce{BA}}_{\ce{BB}} \\ 
\pi^{\ce{AB}}_{\ce{AA}} & \pi^{\ce{AB}}_{\ce{BA}} & 0 & 0 \\ 
0 & 0 & \pi^{\ce{BB}}_{\ce{AB}} & \pi^{\ce{BB}}_{\ce{BB}}  \end{pmatrix}  \nonumber  \\
1 &=& \pi_{\ce{XY}}^{\ce{YA}} + \pi_{\ce{XY}}^{\ce{YB}} 
\eea
Since each 2-monomer state retains its last monomer in its new tip state $\ce{XY} \rightarrow \ce{YZ}$, each tip state can only transition to $s$ tip states in one step, and coefficients for remaining transitions are 0. In general, each of $s^k$ tip states can have $s$ possible transitions, each row in $T_{k,s}$  will thus have $s^{k}-s$ zero entries for $k>0$, and thus becomes exponentially more sparse with $k$. 

One framework for the thermodynamic description considers mass-balance equations which recursively calculate mean-field concentrations of polymer species through elongation equilibria
\bea
\! \! \! \! \! \! \frac{[\ce{X_nXYZ}]_{eq}}{[\ce{X_nXY}]_{eq}} = K_{\ce{XYZ}} [\ce{Z}]_{eq} = \bar{K}_{\ce{XYZ}}
\eea
which are collected in the expression
\bea
\ [\pmb{\ce{X_{n+1}\omega}}] = M_{k,s} [\pmb{\ce{X_{n}\omega}}]
\eea
e.g. for $k=s=2$ we have
\bea
\! \! \! \! \! \!  M_{2,2} &=& \begin{pmatrix}
\bar{K}_{\ce{AAA}} & \bar{K}_{\ce{BAA}} & 0 & 0 \\ 
0 & 0 & \bar{K}_{\ce{ABA}}  & \bar{K}_{\ce{BBA}}  \\ \bar{K}_{\ce{AAB}}  & \bar{K}_{\ce{BAB}} & 0 & 0 \\ 
0 & 0 & \bar{K}_{\ce{ABB}} & \bar{K}_{\ce{BBB}} \\   \end{pmatrix} , \nonumber  
\eea
and, omitting the equilibrium suffix for brevity
\bea
\! \! \! \! \! \! \! \! \! \! \! \! [\pmb{\ce{X_{n}\omega}}] = ([\ce{X_nAA}], [\ce{X_nBA}], [\ce{X_nBA}], [\ce{X_nBA}] )^T .\nonumber 
\eea
The treatments of thermodynamic equilibrium and irreversible elongation have equations of the same functional form, which will allow us to draw several conclusions that apply for equilibrium systems as well as far from equilibrium systems.

%Transitions are probabilities and sum to 1:  %$\pi_{\ce{XY}}^{\ce{YA}} + \pi_{\ce{XY}}^{\ce{YB}} = 1$

\subsection{Periodic sequences}

\begin{figure}[tbhp!]
\centering
\includegraphics[width=1.0\linewidth]{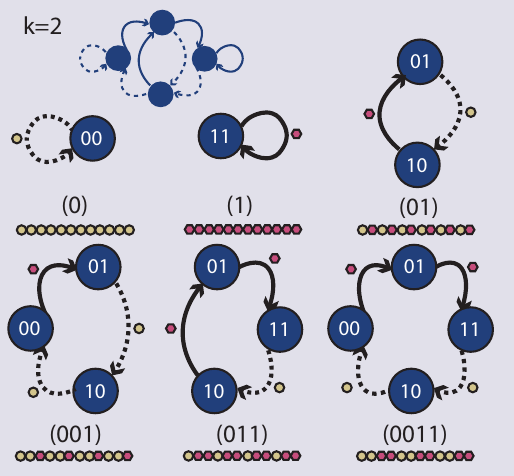}
\caption{Elementary circuits in binary de Bruijn graph for $k=2$. Each elementary circuit can be represented as a subgraph, and corresponds to one of $\gamma=6$ periodic copolymers. A circuit of $r$ reactions corresponds to a Lyndon word of length $r$.}
\label{fig:bruijncircuits}
\end{figure}

A periodic copolymer has a monomer sequence that repeats. For such a sequence to deterministically arise, we require selectively favored reactions. For instance, to obtain ABABAB..., we require that $\ce{PA}$ strongly prefers incorporating $\ce{B}$ over $\ce{A}$ and that $\ce{PB}$ prefers incorporating $\ce{A}$ over $\ce{B}$. A limiting case is where these are the only reactions 
\bea
\ce{PA} + \ce{B} \underset{2}{\leftrightarrows} \ce{PAB} \ \ \ \ \ 
\ce{PB} + \ce{A} \underset{3}{\leftrightarrows} \ce{PBA} \nonumber
\eea
We may denote a repeating unit in brackets (..). The three possible periodic solutions for $k=1,s=2$ can then be represented as ($\ce{A}$), ($\ce{B}$), ($\ce{AB}$), 
\bea
(\ce{A}) &\equiv& \ce{AAAA}...\ , \ \ \  (\ce{B}) \equiv \ce{BBBB}... \nonumber \\
(\ce{AB}) &\equiv& \ce{ABABAB}...\ 
\eea
We may furthermore remark that each periodic sequence corresponds to an elementary circuit of the corresponding de Bruijn Graph, and vice versa Fig.\ref{fig:bruijncircuits}.

Dimonomeric ($s=2$) sequences either come in complementary pairs or are self-complementary, e.g. in Fig \ref{fig:bruijncircuits} the pairs ($\ce{A}$), ($\ce{B}$) and $(\ce{AAB})$, $(\ce{ABB})$ represent the "same" sequence if we exchange the labels $\ce{A} \leftrightarrow \ce{B}$. $(\ce{AB})$ and  $(\ce{AABB})$ come by themselves. For our applications (e.g. determining $\#$ of distinct experimental outcomes), we will count such sequences with repetition. We note that the counting by Balaban and Artemy considers 'reduced sequences' where repetitions are not counted\cite{Balaban1987_I,Artemi1987_II,Balaban1988} (i.e. Fig. \ref{fig:bruijncircuits} contains 4 reduced sequences). 

\subsection{Sequence notation}

We may uniquely represent a repeating sequence through a lexicographical ordering, which functions similar to priority rules in IUPAC nomenclature. We assign some order to successive monomer types $\ce{A}<\ce{B}<..$ and choosing the arrangement in which the "lowest" monomers appear first i.e. $\ce{ABB}<\ce{BAB}<\ce{BBA}$. If we replace monomers $\ce{A},\ce{B},..$ with numbers 0,1,.. the unique representation is then the smallest number, e.g. $011<101<110$.  %so by this ordering we can choose ($011$) to uniquely represent this sequence. 

Such a sequence\cite{Shirshov1953,Lyndon1954_words} that is smaller (lower in lexicographic order) than its rotations is commonly referred to as a \pmb{Lyndon word}\footnote{after Roger Lyndon who researched such words in 1954, and referred to them as 'standard lexicographic sequences'\cite{Lyndon1954_words} They also appeared in 1953 in the work of Anatoly Shirshov under the name 'regular word'\cite{Shirshov1953}.} Sequences obtained as function of $k$ are given in table \ref{tab:tableseq} (see Appendix for higher k sequences).

\subsection{Elementary properties}

We want to consider periodic sequences that on average repeat several times. This requirement gives rise to several properties.

Firstly, periodic repetition requires \textbf{specificity}: each tip state in a periodic sequence must be dominantly followed by a single subsequent tip state. I.e. in ($\ce{AAB}$) $\ce{PAA}$ should preferably add B, $\ce{PAB}$ should preferably add $\ce{A}$, $\ce{PBA}$ should preferably add $\ce{A}$.  

From \textbf{specificity}, we obtain a \textbf{single-repeat law}: in a periodic sequence in k-polymerization, each k-subsequence appears at most once\footnote{since each k-subsequence $\omega$ is selectively followed up by a single k-subsequence $\omega'$, a second occurrence of $\omega$ will be followed up by the same subsequences, and must thus constitute a repeat of the smaller repeating unit.}. 

\textbf{specificity length}: by the same token, there are as many specific reactions* as there are k-tip states, and thus as many specific reactions as the repeating unit is long. We will see later that this specificity can be relaxed for many k-sequences, when some steps are taken to depend on fewer tip monomers.

\textbf{size limit}: A k-Lyndon word can at most be of length $s^k$, by including each of $s^k$ tip states once, and such a cyclic sequence is  called a de Bruijn sequence. For instance for $k=s=2$, there is one de Bruijn sequence, namely ($\ce{AABB}$) containing $\ce{AA, AB, BB, BA}$.

It can be arduous to find all sequences admitted by k-polymerization, it is easy to check a single sequence: any sequence no longer than $s^k$ and obeying the single-repeat law is a sequence that can form by k-polymerization.

\subsection{Finding sequences}

To find all sequences as function of $k,s$, we consider two methods shortly described below. A more elaborate discussion can be found in the SM and accompanying our code on Github.

I) \textbf{single-repeat law filtering} An inefficient method consists in first producing all Lyndon words (e.g. by Duval's  algorithm) up to length $s^k$, and subsequently eliminating those that are not k-Lyndon words, (incompatible with k-polymerization), which are those words that conflict with the single-repeat law. 

For instance, for $k=2, s=2$, we would generate $8$ pre-sorted Lyndon words\footnote{we do not consider the empty Lyndon word of length 0} up to length $4$ :
(A), (B), (AB), (AAB), (ABB), (AAAB), (AABB), (ABBB).

Counting repeats of tips states, we then find that (AAAB) contains AA twice, and (ABBB) contains BB twice and are eliminated.

(A), (B), (AB), (AAB), (ABB), (AABB).

Since the number of Lyndon words grows considerably faster than the number of k-Lyndon words, this method can only be used for small $k,s$.

II) \textbf{simple directed cycles of De Bruijn graph} A more efficient method consists in taking the associated de Bruijn graph and directly calculating the number of simple directed cycles for them. Algorithms for the efficient counting thereof (in our case, Johnson's algorithm) have been implemented in the networkx package.

The list of identified cycles is in turn converted to a list of associated k-Lyndon words, and sorted lexicographically.

\begin{table}[]
    \centering
    \begin{tabular}{|c|c|}
    \hline
        $k$ & Sequences ($s=2$) \\ \hline
        $0$ &  $\textbf{(\ce{A}), \ (\ce{B})}$ \\ \hline 
        $1$ & $(\ce{A}), \ (\ce{B}), \ \textbf{(\ce{AB})}$ \\ \hline
        $2$ & $(\ce{A}), \ (\ce{B}), \ (\ce{AB})$, \ $\textbf{(\ce{AAB}), \ (\ce{ABB}), \ (\ce{AABB})}$  \\ \hline 
        $3$ & $(\ce{A}), \ (\ce{B}), \ (\ce{AB})$, \ $(\ce{AAB}), \ (\ce{ABB}), \ \textbf{(\ce{AAAB})}$  \\ 
         & $(\ce{AABB}), \ \textbf{(\ce{ABBB}), \ (\ce{AAABB})},$ \\
         & $ \textbf{(\ce{AAABB}), \ (\ce{AAABBB}), \ (\ce{AABABB})}$ \\
         &  $\textbf{(\ce{AABBAB}), (\ce{AAABABB}), \ (\ce{AAABBAB})}$ \\
         & $ \textbf{(\ce{AABABBB}), (\ce{AABBBAB})},$  \\ 
         & $ \textbf{ (\ce{AAABABBB}), \  (\ce{AAABBBAB})}$ \\ \hline 
         $k$ & Sequences ($s=3$) \\ \hline
         $0$ &  $\textbf{(\ce{A}), \ (\ce{B}), \ (\ce{C})}$ \\ \hline 
         $1$ &  $(\ce{A}), \ (\ce{B}), \ (\ce{C})$, $\textbf{(\ce{AB}), (\ce{AC})  \ (\ce{BC})}$ \\
          & $ \textbf{(\ce{ABC}), \ (\ce{ACB})}$ \\ \hline 
    \end{tabular}
    \caption{Sequences (Lyndon words) involving $s=2$ monomers as function of vicinality $k$. At its first occurrence (lowest k) each sequence is writen in \textbf{bold}. As the number of sequences grows rapidly with $k,s$ (Table \ref{tab:placeholder}), further sequences are provided in the appendix and SM.}
    \label{tab:tableseq}
\end{table}

\subsection{Sequence diversity}

\begin{figure}[tbhp!]
\centering
\includegraphics[width=1.0\linewidth]{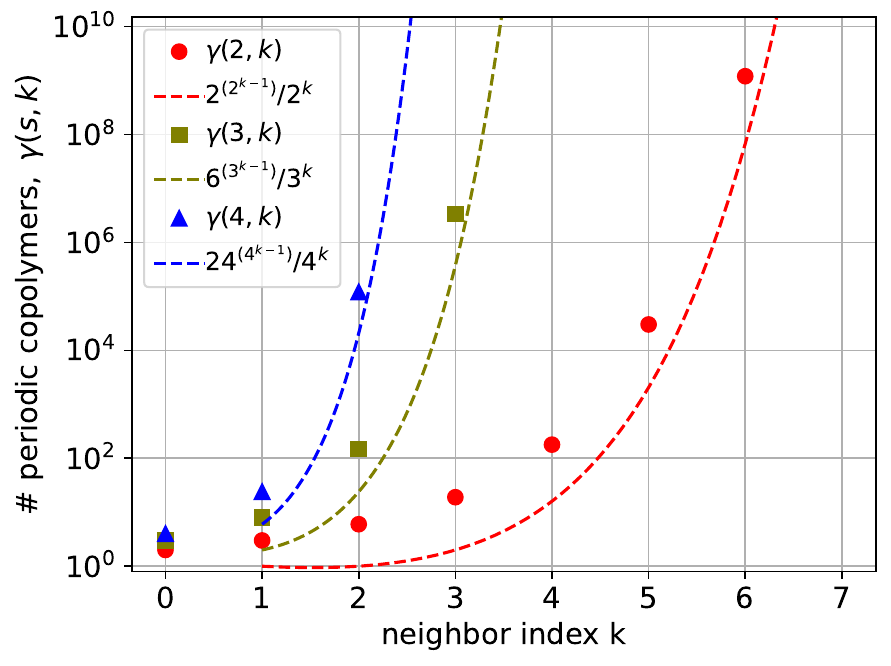}
\caption{Number of periodic copolymers $\gamma(s,k)$ as function of neighbor index $k$ for $s=2,3$ and $4$ monomers. The number of patterns grows explosively with $k$ and $s$. Dotted lines correspond to a lower bound given by the number of DeBruijn sequences of length $s^k$.}
\label{fig:copol}
\end{figure}

As we increase the neighbor index $k$, the number of possible periodic sequences $\gamma$ increases explosively. We can show that $\gamma \approx c^{s^k}$, so that $\gamma(s,k+1) \approx \gamma(s,k)^s$. For $s=2$, increasing neighbor index $k$ by 1 approximately leads to a quadratic increase in periodic sequences. This point is also illustrated in Fig. \ref{fig:copol}

\begin{table}[]
    \centering
    \begin{tabular}{|c|c|c|c|c|c|c|c|}
    \hline
        $k$ &  0 & 1 & 2 & 3 & 4 & 5 & 6  \\ \hline
        $\gamma(2,k)$ & 2 & 3 & 6 & 19 & 179 & 30176 & 1202267287 \\ \hline 
        $\gamma(3,k)$ & 3 & 8 & 148 & 3382522 & . & . & . \\ \hline
        $\gamma(4,k)$ & 4 & 24 & 120538 & . & . & . & . \\ \hline
    \end{tabular}
    \caption{Number of periodic sequences $\gamma(s,k)$, as determined by counting simple directed cycles using Johnson's algorithm. $\gamma(2,k)$ corresponds to A306522 in the OEIS\footnote{https://oeis.org/A306522}. Entries with $\cdot$ were beyond computational reach by our methods.}
    \label{tab:placeholder}
\end{table}

\subsection{Defects and Blocks} 
At each step, copolymerization allows for incorporation of any one among $s$ monomer types. Along repetitions of a periodic copolymer sequence, we thus eventually expect an interruption by incorporation of a defect\footnote{We remark that the occurrence of defects is a generic feature in the chemistry of phases and pseudophases. A phase whose defects are sufficiently negligible and stoichiometry sufficiently fixed is sometimes referred to as a line phase.}, i.e. a monomer different from that dictated by the sequence. For instance, let us consider $k=1$ with favorable incorporation of $\ce{A}$ both for $\ce{PA}$ and $\ce{PB}$. We then expect a copolymer with occasional defects typically of length $1$ .. \ce{A_n B A_m B} .. . Conversely, if $\ce{PB}$ favors incorporation of $\ce{B}$, we obtain a block copolymer with each block corresponding to a periodic sequence $\ce{A_n B_m A_j B_i} .. $.

\begin{figure}[tbhp!]
\centering
\includegraphics[width=1.0\linewidth]{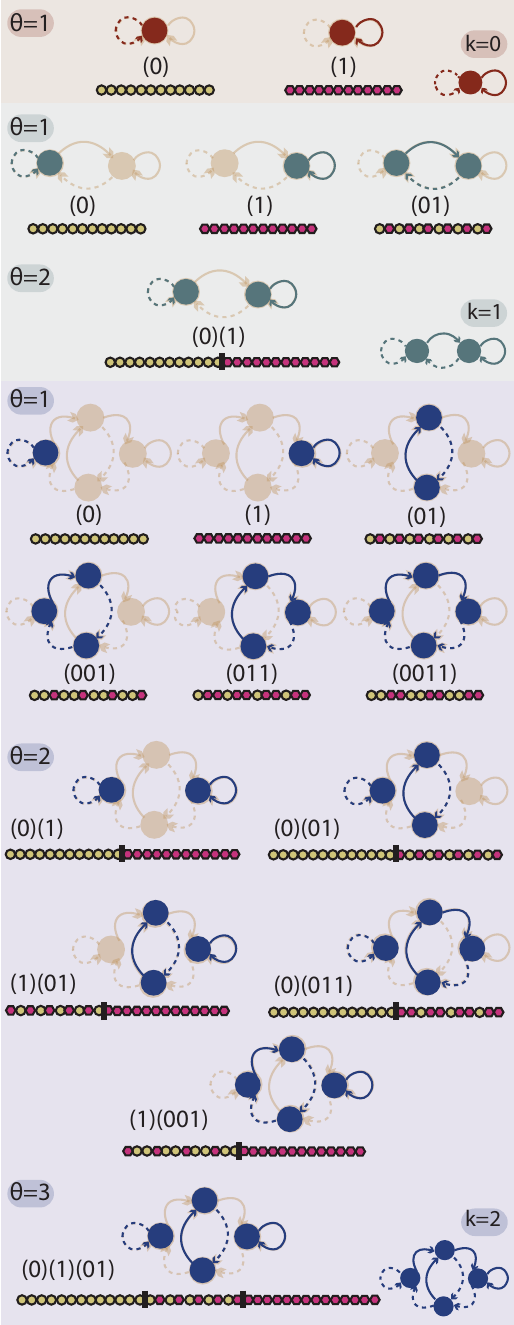}
\caption{Elementary circuits and unions of disjoint elementary circuits for $k=0,1,2$ and $s=2$, corresponding to (periodic) block copolymer solutions with $\theta$ the number of block types.}
\label{fig:block_copo_theta}
\end{figure}

From \textbf{specificity}, it follows that each specific reaction can be present in at most one periodic block. Admissible combinations of periodic blocks are accordingly bound by the same collective size limit $s^k$. Let us denote $\theta$ the number of distinct (periodic) block types in a copolymer. 
For $k=0$, $\theta =1$ is the highest number of periodic blocks. For $k=1$, there is one $\theta=2$ realization $(\ce{A})(\ce{B})$. For $k=2$, we have 5 $\theta = 2$ realizations $(\ce{A})(\ce{B}), (\ce{A})(\ce{AB}), (\ce{A})(\ce{ABB}), (\ce{B})(\ce{AB}), (\ce{B})(\ce{ABB}) $, and a single $\theta=3$ realization $(\ce{A})(\ce{AB})(\ce{B})$. Visually, these correspond to disjoint elementary circuits in the regular de Bruijn graph (Fig. \ref{fig:bruijncircuits}).

\subsection{Thermodynamic vs kinetic block coexistence}

It may be remarked that when blocks become long, it is unlikely to obtain $\theta > s$ at thermodynamic equilibrium. The reason for this is that formally, we can make at most $s$ sequences that are compositionally noninterconvertible, e.g. (A), (B) and any further sequence is interconvertible, i.e. 
\be
...(\ce{A})_n...(\ce{B})_n... \leftrightarrows ...(\ce{AB})_n
\ee
for which the standard free energy change can be considered to be an affine function of $n$
\be
\Delta_r^{\circ} G = g^* + n \Delta g^\circ
\ee
For a fixed per-monomer free energy change $\Delta g^\circ \neq 0$, this equilibrium becomes strongly one-sided as $n$ is increased. Accordingly, as blocks become long enough we expect all such reactions to proceed to completion until at most $s$ blocks remain: $\theta \leq s$.  

In essence, blocks behave as (psuedo)phases (P) and monomers as components (C): when blocks are long enough they will exclude coexistence in a manner that directly analogous to the Gibbs Phase rule\cite{Gibbs1875a} and more specifically the mineralogical phase rule ($P_{eq} \leq C$). Further development of this point\cite{aggnorm2025} is beyond the current scope and will be treated in a separate work.

It may be remarked that an analogous argument does not apply to kinetic copolymerization, i.e. block copolymer $(\ce{A})(\ce{AB})(\ce{B})$ can be thermodynamically excluded (for large blocks), but not kinetically excluded. 

Having established some fundamentals of k-polymerization, it is instructive to now consider how mechanisms with different $k$ emerge, by looking at a simple set of rules applicable to a large body of work in supramolecular and coordination chemistry, and the k-polymerization they can (and cannot) give rise to.

%Within the context of our discussion, 
%To consider a sequence periodic, we require that such a sequence on average repeats several times before a defect may occur. This will in turn require that the sequence is sufficiently favored thermodynamically or that there is sufficient kinetic selectivity towards its formation. We remark that the occurrence of defects is a generic feature in the chemistry of phases and pseudophases. A phase whose defects are sufficiently negligible and stoichiometry sufficiently fixed is sometimes referred to as a line phase.

\section{Simple assembly rules (SAR)}

\begin{figure}[tbhp!]
\centering
\includegraphics[width=1.0\linewidth]{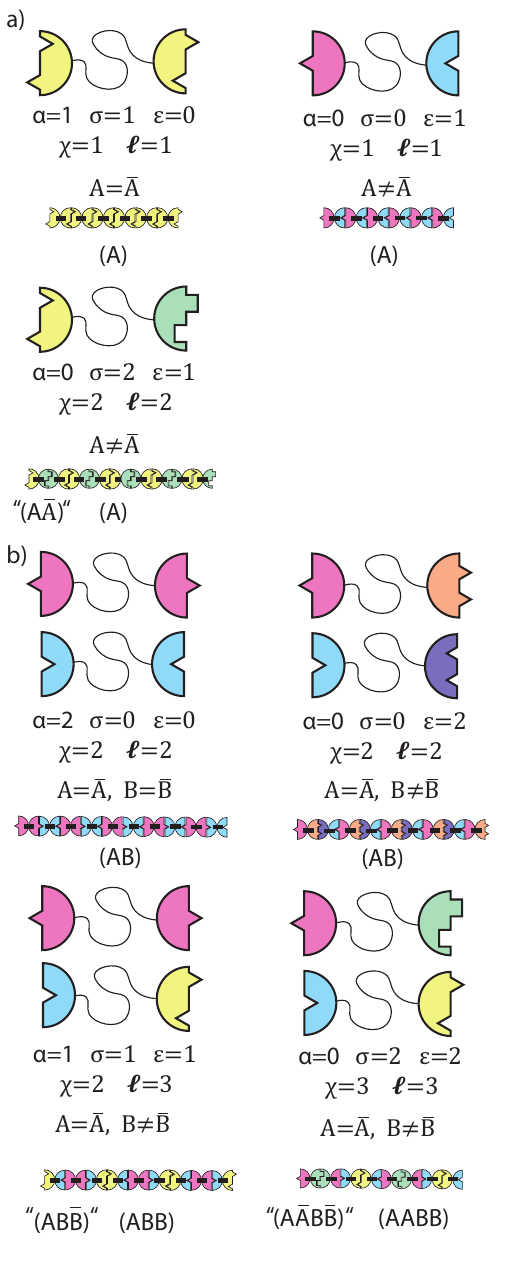}
\caption{Examples of sequences under SAR and associated indices. a) single monomer: I) ($\ce{A}$). II) ($\ce{A}$) III) ($\ce{A}$), from ($\ce{A}\bar{\ce{A}}$). b) two monomers,  IV) ($\ce{AB}$) , V) ($\ce{AB}$), VI) ($\ce{ABB}$), VII ($\ce{AABB}$). All sequences for $s=2,k=2$ can be attained.}
\label{fig:SAR}
\end{figure}

We will now consider a simple set of rules - Simple Assembly Rules (SAR) descriptive of a class of supramolecular chemistry (e.g. involving orthogonal interactions). The rules are \\
\textbf{ditopicity}: a monomer $\ce{A}$ can be represented as two 'sites' $\ce{X},\ce{Y}$: $\ce{A} = \ce{X}-\ce{Y}$ (Fig. \ref{fig:SAR}). A site associates with another site (e.g. $\ce{X}-\ce{Y} \cdot \cdot \ \ce{Z}-\ce{W}$). When sites are distinct ($\ce{X} \neq \ce{Y}$) $\ce{A}$ is  \textit{heteroditopic}, otherwise homoditopic ($\ce{A} = \ce{X}-\ce{X}$). \footnote{we remark that the nature of the sites depends on context. Moreover, the heteroditopic orientation may arise by the assembly process: the original monomer could be symmetric but once assembled two distinct means of assembly may arise.} \\
\textbf{specificity}: each site specifically coordinates with one other site. When a site coordinates with itself $\ce{X-Y} .. \ce{Y-X}$, it is called \textit{self-complementary}. Different monomers do not share the same sites. \\
\textbf{locality}: association depends only on the neighboring binding site (SAR assumes an absence of remote effects) \\
\textbf{2-orientability}: heteroditopic $\ce{A} = \ce{X-Y}$ can be 'flipped around': $\bar{\ce{A}}=\ce{Y-X} \neq \ce{A}$. For homoditopic monomers, $\ce{A} = \bar{\ce{A}}$ ($\ce{X-X}=\ce{X-X}$), whereas for heteroditopic monomers $\ce{A} \neq \bar{\ce{A}}$ ($\ce{X-Y}=\ce{Y-X}$). Site interactions are assumed orientation-free: if site $\ce{Y}, \ce{Z}$ associate in one sense $\ce{X}-\ce{Y} \cdot \cdot \ \ce{Z}-\ce{W}$, they also do in reverse $\ce{W}-\ce{Z} \cdot \cdot \ \ce{Y}-\ce{X}$.

We introduce the following indices: \\ 
$\alpha$: $\#$ distinct homoditopic monomers \\
$\epsilon$ : $\#$ distinct heteroditopic monomer  \\
$\sigma$ : $\#$ distinct self-complementary sites \\
$\chi$ : $\#$ distinct interactions \\
$\ell$ : pattern length \\

To better capture 2-orientability, we first extend the alphabet.

\subsection{Oriented monomer alphabet}

Through the $\textbf{locality}$ property, SAR behavior can be described by $k=1$ k-polymerization with $2s$ oriented monomers ($\ce{A},\bar{\ce{A}},\ce{B},\bar{\ce{B}}, ..$) with the additional property of a \textbf{dyad constraint}: if $\ce{A}$ specifically adds to $\ce{B}$, then $\bar{\ce{B}}$ specifically adds to $\bar{\ce{A}}$. 

In chemistry, sequences are often denoted without this orientational information. We can see (Fig. \ref{fig:SAR}) how this endows SAR assembly with some $k=2$ character: $\ce{A}$ and $\ce{\bar{A}}$ may undergo distinct specific reactions. If this distinction is obscured in the notation, we may determine the orientation of $\ce{A}$ from the preceding monomer. In this way, $(\ce{A}\bar{\ce{A}}\ce{B}\bar{\ce{B}})$ involves $k \geq 1$, whereas $(\ce{AABB})$ requires $k \geq 2$. 

SAR describes one of the mechanisms - 'orientation' - by which higher-k sequences become possible. For $s=2$ monomers, all $k=2$ sequences are mechanistically accessible under $k=1$ SAR. For $s=3$, only a subset of the $k=2$ sequences are accessible, due to the dyad constraint which for instance disallows the sequence $(\ce{ABBC})$.

%we may need to know the identity of a preceding monomer to determine if $\ce{A}$ refers to $\bar{\ce{A}}$ or $\ce{A}$. Similarly, $(\ce{A}\bar{\ce{A}}\ce{B}\bar{\ce{B}})$ can be interpreted as a $k=1$ sequence, whereas $(\ce{AABB})$ requires $k=2$.
\subsection{Example: $\ce{AAB}$}

%Let us treat heteroditopics $\ce{A}, \bar{\ce{A}}$ - distinct orientations of the same monomer - as distinct monomers. By specificity, each tip monomer has a distinct tip site, which in turn binds one 'tail' site specific to one monomer. 
Let us consider ($\ce{A}\bar{\ce{A}}\ce{B}$), with $\ce{A} = \ce{X-Y}$, $\ce{B} = \ce{Z-Z}$, where $\ce{Y}$ binds itself, $\ce{X}$ binds $\ce{Z}$, so $(\alpha,\epsilon,\sigma) = (1, 1, 1)$. The $k=1$ scheme is then 
\bea
\ce{PA} + \bar{\ce{A}} \underset{1}{\leftrightarrows} \ce{PA\bar{A}} \nonumber \\
\ce{P}\bar{\ce{A}} + \ce{B} \underset{2}{\leftrightarrows}  \ce{P\bar{\ce{A}}B} \nonumber \\
\ce{PB} + \ce{A} \underset{3}{\leftrightarrows}  \ce{PBA} \nonumber 
\eea
To describe the same without orientational information, we instead include a second neighbor ($k=2$)
\bea
\ce{PBA} + \ce{A} \underset{1}{\leftrightarrows} \ce{PBAA} \nonumber \\
\ce{PAA} + \ce{B} \underset{2}{\leftrightarrows} \ce{PAAB} \nonumber \\
\ce{PAB} + \ce{A} \underset{3}{\leftrightarrows}  \ce{PABA} \nonumber 
\eea
As there is one tip state with terminal $\ce{B}$, we could also have retained it as $k=1$ reaction
\bea
\ce{PBA} + \ce{A} \underset{1}{\leftrightarrows} \ce{PBAA} \nonumber \\
\ce{PAA} + \ce{B} \underset{2}{\leftrightarrows} \ce{PAAB} \nonumber \\
\ce{PB} + \ce{A} \underset{3}{\leftrightarrows} \ce{PBA} \nonumber 
\eea
More generally, it is only for de Bruijn sequences that all reactions need to be of vicinality k.

\subsection{SAR sequence space}
\label{subsec:SARseq}
In the SM we derive the SAR sequence space, which follows from the dyad constraint and is captured by two patterns: \\
\textbf{asymmetric} ($\alpha=\sigma=0) \ \ (\ce{AB}...\ce{MN})$ \\
\textbf{symmetric} $(\alpha + \sigma=2) \ \  (\square \ce{B}...\ce{M}\Diamond \bar{\ce{M}}...\bar{\ce{B}})$ \\
where \ \ $\square = \ce{A} \lor \ce{A}\bar{\ce{A}}, \ \  $  $\Diamond = \emptyset \lor \ce{N} \lor \ce{N}\bar{\ce{N}}$  

It may be observed that all but one sequence ($(\ce{A}\bar{\ce{A}}) \rightarrow (\ce{A})$) are unaltered by going to an orientation-free alphabet. 

Note that this include all patterns for $k=2,s=2$, (A),(B),(AB),(AAB),(ABB),(AABB), where (A), (B), (AB) can be either symmetric or asymmetric. (AAB), (ABB) and (AABB) can only be symmetric under SAR. This and some other behavior can be captured by articulating a few SAR laws.

\subsection{SAR laws}

\textbf{Reversal law}: a sequence either has two elements of reversal (homoditopics, self-complementary sites), or none
%: either a sequence is asymmetric, or it is symmetric, in which case a pattern must have two elements of reversal symmetry. Such elements either come from homoditopics or self-complementary sites.
\be
\alpha + \sigma = 2 \lor 0
\ee 
\textbf{Length-symmetry law}: asymmetric sequences are purely heteroditopic, each monomer occurs once. In symmetric sequences heteroditopics occur twice, homoditopics once.
\bea
\ell = \epsilon + \frac{\alpha + \sigma}{2} \left(\epsilon + \alpha \right) 
\eea
Where the length is that of the motif in which $\ce{X}, \bar{\ce{X}}$ are treated as distinct. This coincides with the length of the orientation-free sequence, except for $(\ce{X} \bar{\ce{X}})$ (see Sec.\ref{subsec:SARseq}), which would collapse to $(\ce{X})$.

%of the $2s-$letter motif where $\ce{X}, \bar{\ce{X}}$ are treated as distinct. This coincides with the $s-$letter length, except for $\ce{A}\bar{\ce{A}}$.

\textbf{Interaction-symmetry law}: asymmetric sequences have as many distinct interactions as there are heteroditopics. Symmetry and self-complementarity each add 1 interaction.
\bea
%\chi &=& \epsilon \ \ \ (\alpha + \sigma = 0) \\
%\chi &=& 1 + \epsilon + \sigma \  \ \ (\alpha + \sigma = 2)  \\
\chi = \epsilon + \frac{\alpha + \sigma}{2} \left(1 + \sigma \right)
\eea
We can combine these laws to characterize the number of distinct interactions needed to attain a desired sequence of length $\ell$ \\

\textbf{Length-interaction-symmetry law}: in an asymmetric sequence the pattern length equals the $\#$ distinct interactions. With symmetry, the $\#$ interactions decreases by 1 for each distinct monomer, increases by 1 per distinct self-complementary interaction and 1 overall.
\be
\ell = \chi +  \frac{\alpha + \sigma}{2} \left(\epsilon + \alpha - 1 - \sigma \right)
\ee
Larger patterns thus require fewer distinct interactions when the pattern is symmetric.

%\section{Forbidden patterns}

%\ce{AB\bar{A}\bar{B}}

\subsection{Transcending SAR: accessing complex sequences}

%Through our characterization of SAR, we have seen what sequences can and cannot be obtained from a set of commonly applicable basic rules. 
%The SAR model provides us with a methodology towards designing complex sequences: 

To obtain sequences beyond SAR, we require designs that circumvent at least one SAR rule. A particularly restrictive rule is locality, which rather directly enforces $k=1$. 

One natural strategy to circumvent this locality is to have long-range interactions, which may be repulsive or attractive. One repulsive strategy is the introduction of steric hindrance that enforce a minimal distance among specific monomer types. We found a literature report of a complex $k=4$ sequence ($\ce{AAAABB}$), which cannot form under SAR\cite{gorl2015}. This system involves a sterically crowded monomer B that favorably forms dimers with itself, but - due to steric hindrance - not trimers. Bulky $\ce{BB}$ dimers copolymerize favorably with uncrowded monomer $\ce{A}$ and seem to require a separation by 4 $\ce{A}$ monomers to incorporate.

Similarly, one may imagine long-range attractive interactions, e.g. by incorporating long rigid elements to position functional groups beyond the nearest neighbor. Such interactions may also appear structurally, for instance in the quasi-1D organization of a helix, where a monomer that adds to a tip may also feel the effect of another monomer in the helix situated one pitch below. The 1959 Zimm-Bragg model\cite{Zimm1959} for the helix-to-random coil transition in polypeptides considered this point explicitly, and formulated a transfer matrix for arbitrary $k$, focusing in particular on $k=3$ and $k=1$. Coiled helices with $k=3$ interaction have also been prepared supramolecularly, for instance helical nanotubes formed by N,N’-dimethylnaphthalenediimide with 
pendant carboxylic acid group\cite{Pantos2007_helical_tubes}.

To extend beyond 2-orientability, we may imagine other contributions than orientation that endow monomers (or tip\cite{Coleman1963}, or catalyst\cite{coates1995}) with additional states. As has been seen for 2-orientability, such additional states provide a fruitful mechanism to access sequences otherwise not attainable. Such degrees of freedom can e.g. be different modes of chiral incorporation or conformational isomerism. 

It may be remarked that nontrivial sequences are also expected to be harder to sequence, and more readily interpreted as a simpler sequence. For instance when a stoichiometric ratio A:B (e.g. 1:1, 2:1) is observed, it may be parsimonious\footnote{at least, in the absence of any further evidence to the contrary} to interpret this as evidence for the simplest sequence with that composition (e.g. $\ce{AB}, \ce{AAB}$) rather than a more complex sequence (e.g. $\ce{AABB}$, $\ce{AAAABB}, \ce{AAABAB}$).

%It is conceivable that more complex sequences have been prepared

%For instance, monomer and polymer chain (as well as catalyst) can exhibit chirality, which result in multiple nonequivalent ways to incorporate the same monomer.

\section{Application: orthogonal interactions and supramolecular polymers}

Sufficiently orthogonal interactions can be exploited to realize more complex patterns. 

\begin{figure}
    \centering
    \includegraphics[width=1.0\linewidth]{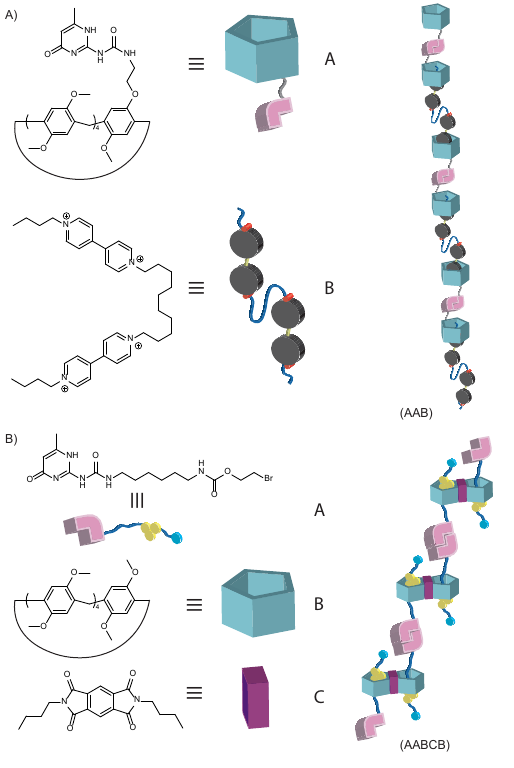}
    \caption{\textbf{Patterns through orthogonal interactions} A) (AAB) through hydrogen bonding and cryptands \cite{Guan2012}. B) (AABCB) through hydrogen bonding, cryptands, pi-stacking.\cite{Wang2017} }
    \label{fig:examplessupra}
\end{figure}

Guan et al \cite{Guan2012} combine A, bearing a self-complementary UPy (L-shape in Fig. \ref{fig:examplessupra}), and a barrel-shaped moiety (pillar[5]arene, drawn as pentagonal barrel) that can host one guest moiety. Here, B contains two bisparaquat groups, which act as a guest that can be encapsulated (drawn as hockeypucks). These motifs induce next nearest-neighbor interaction (k=2): UPy combines with another UPy, but subsequently, no further UPy is available
\bea
\ce{A} &+& \ce{A} \leftrightarrows \ce{AA} \\
\ce{AA} &+& \ce{AAA} \  \xcancel{\leftrightarrows} \ \ce{A_3}
\eea
Similarly, the barrel moiety can host one guest, but not two, preventing $\ce{BAB}$ to form:
\bea
\ce{B} &+& \ce{A} \leftrightarrows \ce{BA} \\
\ce{BA} &+& \ce{B} \ \xcancel{\leftrightarrows} \ \ce{BAB}
\eea
$\ce{B}$ strongly interacts with $\ce{A}$ but not with itself. It follows that $\ce{PAA}, \ce{PAB}, \ce{PBA}$ each incorporate one monomer with high specificity, and eo ipso that we have $3$ specific interactions. Note that two of these interactions are 'the same', but placed in reverse: $\ce{AB}$ and $\ce{BA}$ involve the exact same host-guest interaction, with A bearing a UPY and pillar[5]arene moiety, and B a ditopic species with two bisparaquat groups, which can each be hosted by one pillar[5]arene. Species A dimerizes through a UPY moiety, in turn affording a ditopic species, allowing an (AAB) copolymer to form. Tangxin et al \cite{Tangxin2013} achieved an (AAB) copolymer through self-complementary hydrogen bonding and coordination of secondary ammonium ions with crown ether groups.

Wang et al \cite{Wang2017} prepared an (AABCB) using pillararene, hydrogen bonding and pi-stacking. Another reported\cite{Wang2017b} (AABCB) involved hydrogen bonding, a cryptand with guest, and pillarene with guest. Shi et al prepared (ABCB) \cite{Shi2015}, using terpyridine-zinc coordination, and a pair of heterocomplementary hydrogen-bonding motifs. Shangguan et al \cite{Shangguan2017} prepared an (ABCDCB) motif, using $\ce{Eu3+}$ and N,N’-bis(nbutyl) pyromellitic diimide as homoditopic species.

A racemic mixture of chiral PBIs was found to form (AABB) structures \cite{Wehner2020}. %Although determining sequences in supramolecular polymers is hard, we found one report of a complex sequence (AAAABB), which cannot form under SAR\cite{gorl2015}. This system involves a sterically crowded monomer B that favorably forms dimers with itself, but - due to steric hindrance - not trimers. Dimers copolymerize favorably with uncrowded monomer A, presumably requiring a separation by 4 A.

\begin{table}[]
    \centering
    \begin{tabular}{|c|c|c|c|c|c|c|c|}
    \hline
        Sequence & s & $\alpha$ & $\epsilon$ & $\sigma$  & $\chi$ & $\ell$ & Ref  \\ \hline  
        (A)   &  1    &    0   &    1      &   0     &   1  &  1  &  \cite{Sijbesma1997Asym,Appel2014,Lu2021Asym}
         \\ \hline        
        (A)   &  1    &    0   &    1      &   0     &   1  &  1  &  \cite{Wang2015_heteroditopicsingle,Hirao2019,Hisano2025,Fernandez2008,Zhang2020Ahet,Zhang2011het,Miyauchi2004} %head-to-tail
         \\ \hline        
        (AB)   &  2    &    0   &    2      &   0     &   2  &  2  & \cite{Ogoshi2012,Li2016ABasym,Memis2024,Tomimasu2009,Miyauchi2004,Yamaguchi1999,Memis2025,Niu2011ABsym,Hirao2021ABsym,Hirao2021ABsymx,Tancini2010ABsym} \\ \hline       
        (AB)   &  2    &    2   &    0      &   0     &   1  &  2  & \cite{Blake2000,Yount2003,El-Ghayoury2002,Chen2014,Wang2008AB,Wang2017ABsymphoto,Kimura1999} \\ \hline
        (AAB)   &  2    &    1   &    1      &   1     &   2  &  3  & \cite{Guan2012,Tangxin2013,Wang2013AAB,Yagai2009AAB,Li2023AAB}    \\ \hline 
        (AABB)  &  2    &    0   &    2      &   2     &   3  &  4  &  \cite{Wehner2020}    \\ \hline 
        (ABC)  &  3    &    0   &    3      &   0     &   3  &  3  &  \cite{Hirao2017,Yang2015}    \\ \hline        
        (ABCB)  &  3    &    2   &    1      &   0     &   2  &  4  &  \cite{Shi2015,Zeng2015,Lin2016b,Zhang2022ABCB,Xia2024,Bera2020}    \\ \hline
        (AABCB) &  3    &    0   &    2      &   2     &   3  &  4  &  \cite{Wang2017,Wang2017b}    \\ \hline
        (AABCDCB) &  4   &   0   &    2      &   2     &   3  &  4  &  \cite{Shangguan2017}    \\ \hline
    \end{tabular}
    \caption{Indices for periodic supramolecular copolymer sequences captured by SAR, with selected literature reports of synthetic examples.}
    \label{tab:placeholder}
\end{table}

%\section{Discussion}

\section{Acknowledgements}

A.B. acknowledges the EU (MSCA-PF-2023 ‘KENA’ no. 101155395). This work has been supported by the Novo Nordisk Foundation (grant NNF21OC0066551 ‘MATOMIC’). Research in the Stadler lab is supported by the BMBF (Germany) through DAAD project 57616814 (SECAI, School of Embedded Composite AI).  A.B. acknowledges fruitful conversations with Thomas Hermans and Marten van der Ploeg.

\bibliography{BibFile_final}

\section{Appendix I: Sequence space}

In this section, further k-Lyndon words are provided as function of $k,s$, insofar as this is feasible here. In our repository files can be found containing k-Lyndon words for higher values of $k$. For concision, monomers in this section are represented by successive integers $0,1,..$. 

k-Lyndon words are also visualized in ordered manner in figures  \ref{fig:heatmap}, \ref{fig:heatmapk3s3}, \ref{fig:heatmapk4s2} below. Due to the ordering, there is considerable overlap among the first few positions, which becomes ever thinner as we pass more positions where sequences can differ, until eventually the picture becomes blurry.

\begin{figure}[tbhp!]
\centering
\includegraphics[width=1.0\linewidth]{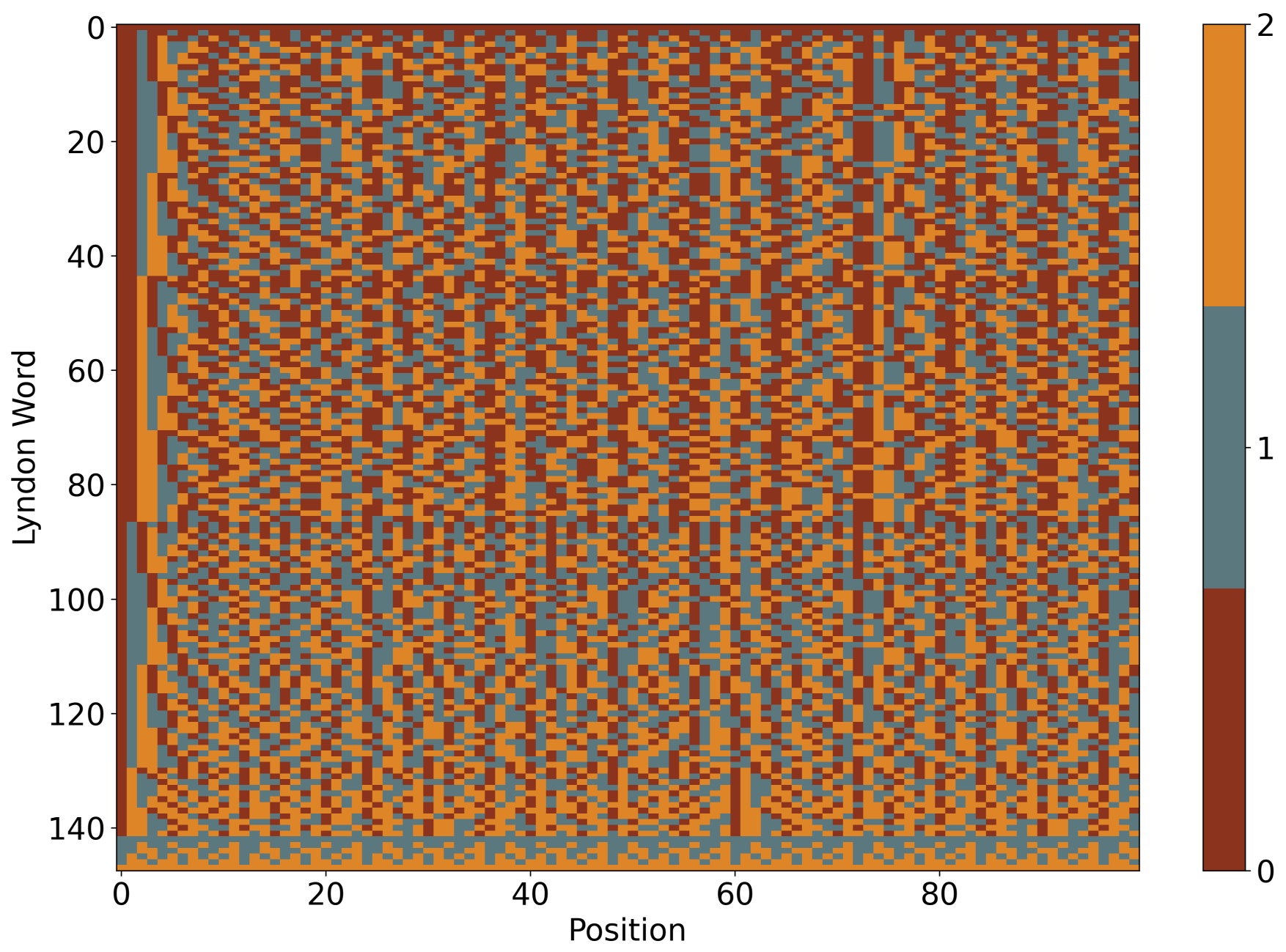}
\caption{A heatmap representation of the sequence space of $k=2,s=3$, $\gamma(2,3)=179$.}
\label{fig:heatmap}
\end{figure}

\begin{figure}[tbhp!]
\centering
\includegraphics[width=1.0\linewidth]{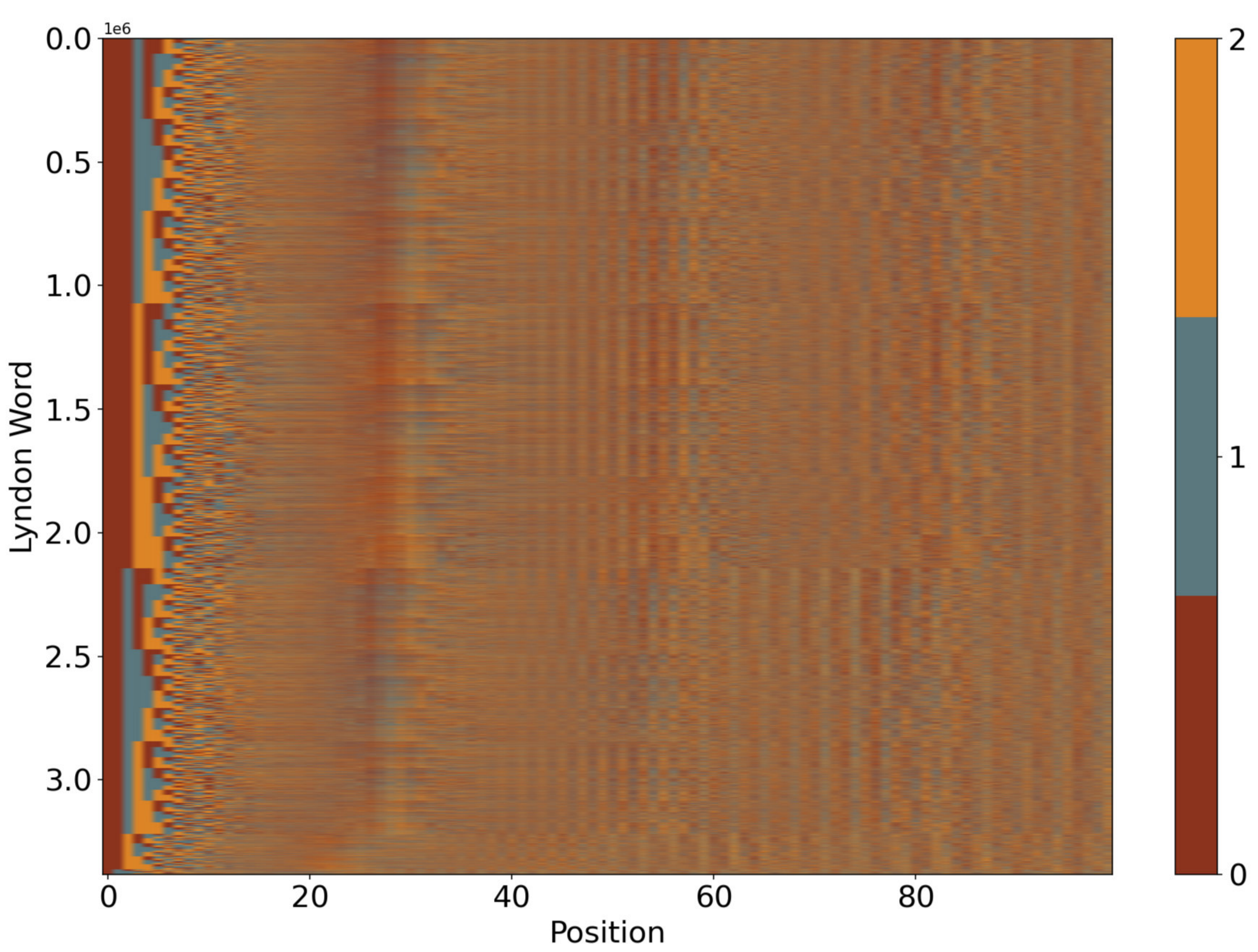}
\caption{A heatmap representation of the sequence space of $k=3,s=3$, $\gamma(3,3)=3382522$.}
\label{fig:heatmapk3s3}
\end{figure}

\begin{figure}[tbhp!]
\centering
\includegraphics[width=1.0\linewidth]{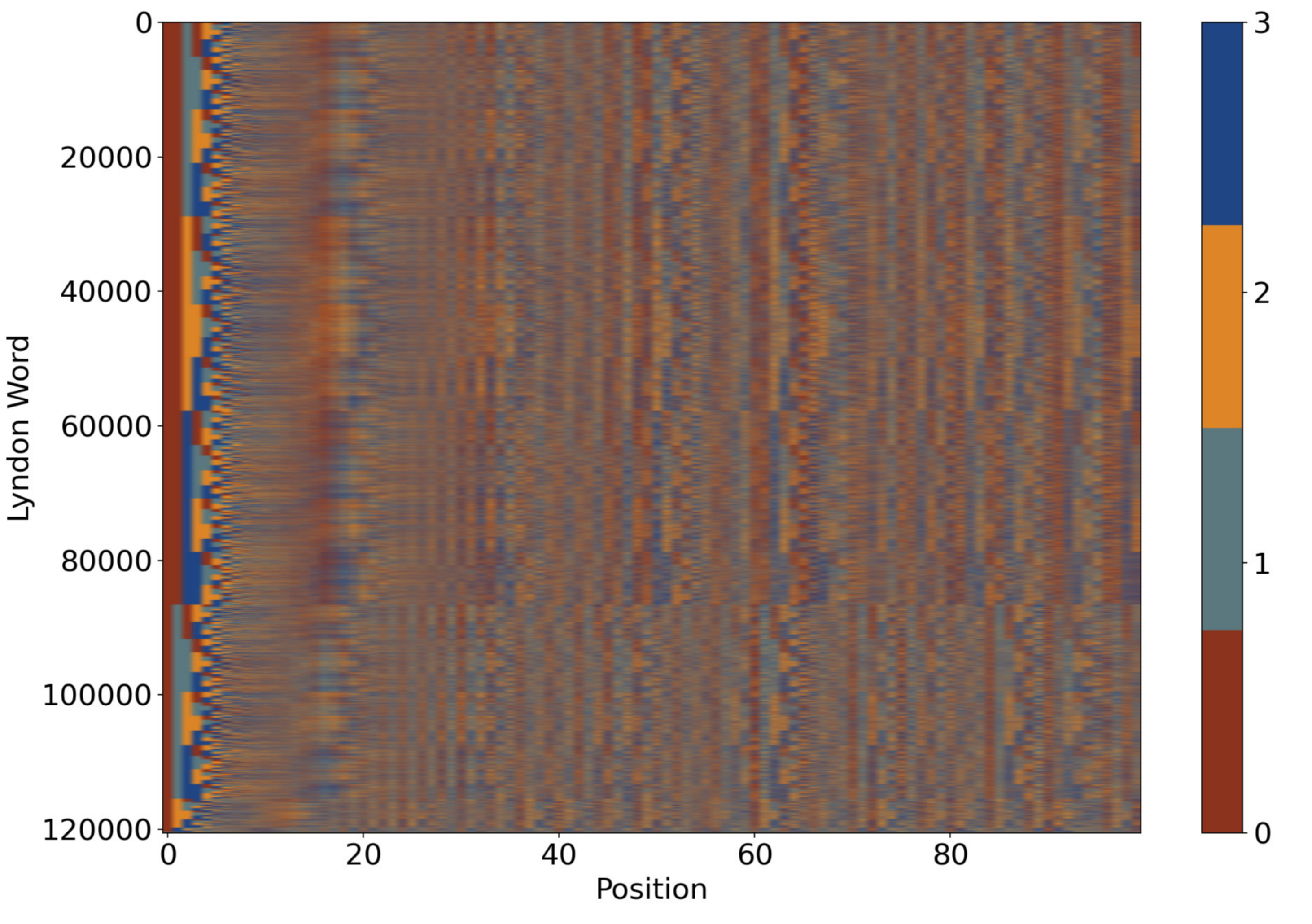}
\caption{A heatmap representation of the sequence space of $k=2,s=4$, $\gamma(4,2)=20538$.}
\label{fig:heatmapk4s2}
\end{figure}

\subsection{k-Lyndon words for k=0, s=2} 

[0, 1]

\subsection{k-Lyndon words for k=1, s=2} 

[0, 1, 01]

\subsection{k-Lyndon words for k=2, s=2} 

[0, 1, 01, 001, 011, 0011]

\subsection{k-Lyndon words for k=3, s=2} 

[0, 1, 01, 001, 011, 0001, 0011, 0111, 00011, 00111, 000111, 001011, 001101, 0001011, 0001101, 0010111, 0011101, 00010111, 00011101]

\subsection{k-Lyndon words for k=4, s=2} 

[0,
 1,
 01,
 001,
 011,
 0001,
 0011,
 0111,
 00001,
 00011,
 00101,
 00111,
 01011,
 01111,
 000011,
 000101,
 000111,
 001011,
 001101,
 001111,
 010111,
 0000101,
 0000111,
 0001011,
 0001101,
 0001111,
 0010111,
 0011101,
 0101111,
 00001011,
 00001101,
 00001111,
 00010011,
 00010111,
 00011001,
 00011101,
 00101101,
 00101111,
 00110111,
 00111011,
 00111101,
 000010011,
 000010111,
 000011001,
 000011101,
 000100111,
 000101101,
 000101111,
 000110111,
 000111001,
 000111011,
 000111101,
 001011101,
 001101111,
 001111011,
 0000100111,
 0000101101,
 0000101111,
 0000110111,
 0000111001,
 0000111011,
 0000111101,
 0001001111,
 0001010011,
 0001011101,
 0001100101,
 0001101111,
 0001111001,
 0001111011,
 0010111101,
 0011010111,
 0011101011,
 00001001111, 00001010011, 00001011101, 00001100101, 00001101111, 00001111001, 00001111011, 00010100111, 00010111101, 00011010111, 00011100101, 00011101011, 00110101111, 00111101011,
 000010100111, 000010111101, 000011010111, 000011100101, 000011101011, 000100110111, 000100111011, 000101001111, 000110101111, 000110111001, 000111011001, 000111100101, 000111101011, 0000100110111, 0000100111011, 0000101001111, 0000110101111, 0000110111001, 0000111011001, 0000111100101, 0000111101011, 0001001101111, 0001001111011, 0001101111001, 0001111011001,
 00001001101111, 00001001111011, 00001101111001, 00001111011001, 00010011010111, 00010011101011, 00010100110111, 00010100111011, 00010110011101, 00010110100111, 00010111001101, 00010111010011, 00011001011101, 00011010010111, 00011010111001, 00011011100101, 00011100101101, 00011101001011, 00011101011001, 00011101100101, 000010011010111, 000010011101011, 000010100110111, 000010100111011, 000010110011101, 000010110100111, 000010111001101, 000010111010011, 000011001011101, 000011010010111, 000011010111001, 000011011100101, 000011100101101, 000011101001011, 000011101011001, 000011101100101, 000100110101111, 000100111101011, 000101001101111, 000101001111011, 000101100111101, 000101101001111, 000101111001101, 000101111010011, 000110010111101, 000110100101111, 000110101111001, 000110111100101, 000111100101101, 000111101001011, 000111101011001, 000111101100101,
 0000100110101111, 0000100111101011, 0000101001101111, 0000101001111011, 0000101100111101, 0000101101001111, 0000101111001101, 0000101111010011, 0000110010111101, 0000110100101111, 0000110101111001, 0000110111100101, 0000111100101101, 0000111101001011, 0000111101011001,  0000111101100101]

\subsection{k-Lyndon words for k=0, s=3} 

[0, 1, 2]

\subsection{k-Lyndon words for k=1, s=3} 

[0, 1, 2, 01, 02, 12, 012, 021] \\

\subsection{k-Lyndon words for k=2, s=3} 

[0, 1,
 2,
 01,
 02,
 12,
 001,
 002,
 011,
 012,
 021,
 022,
 112,
 122,
 0011,
 0012,
 0021,
 0022,
 0102,
 0112,
 0121,
 0122,
 0211,
 0212,
 0221,
 1122,
 00102,
 00112,
 00121,
 00122,
 00201,
 00211,
 00212,
 00221,
 01022,
 01102,
 01121,
 01122,
 01211,
 01221,
 02112,
 02122,
 02211,
 02212,
 001022,
 001102,
 001121,
 001122,
 001211,
 001221,
 002011,
 002112,
 002122,
 002201,
 002211,
 002212,
 010212,
 011022,
 011221,
 012021,
 012102,
 012211,
 021122,
 022112,
 0010212,
 0011022,
 0011221,
 0012021,
 0012102,
 0012211,
 0020121,
 0021012,
 0021122,
 0021201,
 0022011,
 0022112,
 0102112,
 0102122,
 0102212,
 0110212,
 0112021,
 0112102,
 0120211,
 0120221,
 0121022,
 0121102,
 0122021,
 0122102,
 00102112,
 00102122,
 00102212,
 00110212,
 00112021,
 00112102,
 00120211,
 00120221,
 00121022,
 00121102,
 00122021,
 00122102,
 00201121,
 00201211,
 00201221,
 00210112,
 00210122,
 00211012,
 00211201,
 00212011,
 00212201,
 00220121,
 00221012,
 00221201,
 01021122,
 01022112,
 01102122,
 01102212,
 01120221,
 01121022,
 01122021,
 01122102,
 01202211,
 01211022,
 01220211,
 01221102,
 001021122,
 001022112,
 001102122,
 001102212,
 001120221,
 001121022,
 001122021,
 001122102,
 001202211,
 001211022,
 001220211,
 001221102,
 002011221,
 002012211,
 002101122,
 002110122,
 002112201,
 002122011,
 002201121,
 002201211,
 002210112,
 002211012,
 002211201, 
 002212011]

\section{Appendix II: formalizing SAR}

We encode the dominant transitions of each monomer - for both orientations - in a matrix $\partial$, with entries  $\partial_{\ce{X}}^{\ce{Y}} = 0 \lor 1$. 

For an oriented alphabet, we have $k=1$. Letting $s=2$, the matrix $\partial$ is

\bea
\partial = \begin{pmatrix}
    \partial_{\ce{A}}^{\ce{A}} &    \partial_{\bar{\ce{A}}}^{\ce{A}} &    \partial_{\ce{B}}^{\ce{A}} &    \partial_{\bar{\ce{B}}}^{\ce{A}} \\
        \partial_{\ce{A}}^{\bar{\ce{A}}} &    \partial_{\bar{\ce{A}}}^{\bar{\ce{A}}} &   \partial_{\ce{B}}^{\bar{\ce{A}}}   &   \partial_{\bar{\ce{B}}}^{\bar{\ce{A}}} \\
    \partial_{\ce{A}}^{\ce{B}} &    \partial_{\bar{\ce{A}}}^{\ce{B}} &    \partial_{\ce{B}}^{\ce{B}} &    \partial_{\bar{\ce{B}}}^{\ce{B}} \\
        \partial_{\ce{A}}^{\bar{\ce{B}}} &    \partial_{\bar{\ce{A}}}^{\bar{\ce{B}}} &   \partial_{\ce{B}}^{\bar{\ce{B}}}   &   \partial_{\bar{\ce{B}}}^{\bar{\ce{B}}}        
\end{pmatrix} 
\eea
Due to specificity, each monomer tip state has one incoming reaction and one outgoing reaction, i.e. each row and column contain exactly one 1 and zeros elsewhere. 
\bea
\sum_{Y} \partial_{\ce{X}}^{\ce{Y}} = 1 \\
\sum_{X} \partial_{\ce{X}}^{\ce{Y}} = 1 
\eea
i.e. $\partial$ is a permutation matrix. However, it cannot be any permuation matrix: interactions are retained when we flip a dyad $\ce{XY}$ to $\bar{\ce{Y}}\bar{\ce{X}}$, resulting in a constraint for all elements, except self-complementary ones 
\bea
\pi_{\ce{Y}}^{\ce{X}} &=& \pi_{\bar{\ce{X}}}^{\bar{\ce{Y}}} \\ 
\pi_{\ce{X}}^{\bar{\ce{X}}} &=& \pi_{\bar{\bar{\ce{X}}}}^{\bar{\ce{X}}} = \pi_{\ce{X}}^{\bar{\ce{X}}}   
\eea
which restricts the accessible sequences. For instance, $\ce{A}\bar{\ce{A}}\ce{B}\bar{\ce{B}}$ is a possible sequence under SAR, but $\ce{A}\ce{B}\bar{\ce{A}}\bar{\ce{B}}$ is not: if $\ce{B}$ adds to $\ce{A}$ ($\ce{AB}$) then $\bar{\ce{A}}$ adds to $\bar{\ce{B}}$ ($\bar{\ce{B}}\bar{\ce{A}}$). For $\ce{A}\bar{\ce{A}}\ce{B}\bar{\ce{B}}$, $\partial$ becomes
\bea
\partial_{\ce{A}\bar{\ce{A}}\ce{B}\bar{\ce{B}}} = \begin{pmatrix}
    0 &    0 &   0 &    1 \\
        1 &   0 &   0  &   0 \\
    0 &    1 &   0 &   0\\
        0 &    0 &   1   &   0      
\end{pmatrix} 
\eea
The "forbidden" matrix that would - without constraints - give rise to $\ce{A}\ce{B}\bar{\ce{A}}\bar{\ce{B}}$ is 
\bea
\partial^{\times} = \begin{pmatrix}
    0 &    0 &   0 &    1 \\
        0 &   0 &   1  &   0 \\
    0 &    1 &   0 &   0\\
       1 &    0 &   0   &   0      
\end{pmatrix} 
\eea

If a species is homoditopic $\ce{X} = \bar{\ce{X}}$, we can represent it as a single monomer. E.g. letting $\ce{B} = \bar{\ce{B}}$, the $s=2$ matrix can be reduced to

\be
\begin{pmatrix}
    \pi_{\ce{A}}^{\ce{A}} &    \pi_{\bar{\ce{A}}}^{\ce{A}} &    \pi_{\ce{B}}^{\ce{A}}  \\
        \pi_{\ce{A}}^{\bar{\ce{A}}} &    \pi_{\bar{\ce{A}}}^{\bar{\ce{A}}} &   \pi_{\ce{B}}^{\bar{\ce{A}}}   \\
    \pi_{\ce{A}}^{\ce{B}} &    \pi_{\bar{\ce{A}}}^{\ce{B}} &    \pi_{\ce{B}}^{\ce{B}}     
\end{pmatrix}
\ee
for which the dyad constraint still applies, but with $\ce{B} = \bar{\ce{B}}$
\be
\pi_{\ce{B}}^{\ce{A}} = \pi_{\bar{\ce{A}}}^{\ce{B}} 
\ee

\end{document}